\documentclass [aps,pre,showpacs,twocolumn] {revtex4}
\usepackage[final]{graphics}
\usepackage{amssymb}
\usepackage{amsfonts}
\usepackage{epsfig}
\usepackage{color}
\begin{document}
\date{\today}
\title{Adsorption of flexible polymer chains on a surface: Effects of different
solvent conditions}
\author{ P. H. L. Martins$^{1}$, J. A. Plascak$^{2,3}$, and M. Bachmann$^{1,3}$}
\affiliation{$^{1}$Instituto de F\'isica, Universidade Federal de Mato Grosso,
78060-900 Cuiab\'a, MT - Brazil\\
$^{2}$Departamento de F\'isica, Centro de Ci\^encias Exatas e
da Natureza, CCEN, Universidade Federal da Para\'iba, Cidade Universit\'aria,
58051-970 Jo\~ao  Pessoa, PB - Brazil\\
$^{3}$Center for Simulational Physics, University of Georgia, Athens, GA
30602, USA\\
}
\begin{abstract}
Polymer chains undergoing a continuous adsorption-desorption transition  
are studied through extensive computer simulations. A
three-dimensional self-avoiding walk lattice model of a polymer chain grafted
onto a surface has been treated for different solvent conditions. We have used an
advanced contact-density chain-growth algorithm, in  which the density of
contacts can be directly obtained. From this quantity, the order parameter and
its fourth-order Binder cumulant are computed, as well as the corresponding critical
exponents and the adsorption-desorption transition temperature. 
As the number of configurations with a given number of surface contacts 
and monomer-monomer 
contacts is independent of the temperature and solvent conditions,
it can be easily applied to get results for different solvent parameter values 
without the need of any extra simulations. In analogy to continuous 
magnetic phase transitions, finite-size-scaling methods have been employed. 
Quite good results for the critical properties and phase diagram of very long
single polymer chains have been obtained by properly taking into account the
effects of corrections to  scaling. The study covers all solvent effects, going 
from the limit of {\it super-self-avoiding walks}, characterized by
effective monomer-monomer repulsion, to poor solvent conditions that enable the 
formation of compact polymer structures.
\end{abstract}
\maketitle

\section{INTRODUCTION} 

The study of adsorption of polymer chains from a solution onto a flat solid
surface has been extensively investigated for more than 60 years~\cite{sinha},
not only due to its relevance for potential technological and 
biological applications~\cite{gennes,milner,meredith,diaz,bach14}, but also for
its importance on many phenomena such as adhesion, surface coating, wetting,
adsorption chromatography, among others (see, for example, Refs.~\cite{bach14,fleer}). 
In the diluted regime, the chains can be considered
independent of each other, and it is sufficient to investigate the surface
effects on the conformations of a single polymer chain only. Such conformations, in
their turn, can be determined by the temperature of the heat bath, the
corresponding solvent quality, as well as the strengths of the monomer-monomer
and monomer-surface interactions. In general, at sufficiently high temperatures and
good solvent conditions, the polymer chain is expected to be extended and desorbed.
However, at low temperatures, even a small attractive surface interaction is capable of
keeping chain segments adsorbed~\cite{bach14,eisen}. As a result, a continuous
adsorption-desorption (A-D) transition occurs at a critical temperature $T_a$,
with a desorbed phase for $T>T_a$ and an adsorbed phase for $T<T_a$.

An appropriate order parameter for this A-D transition can be given by the ratio
\begin{equation}
n_s= N_s/N, 
\label{ns}
\end{equation}
where $N_s$ is the number of monomers in contact with the surface and
$N$ is the total length of the chain. It is clear that in the desorbed phase
(for $T>T_a$), for very long chains, $n_s\rightarrow0$. Thus,
at the transition temperature $T_a$, a crossover exponent $\phi$ is defined for the
behavior of the mean value of $N_s$ as a function of the chain length by~\cite{eisen}
\begin{equation}
\langle N_s \rangle\sim N^\phi,~~\mbox{~~or}~~~~\langle n_s \rangle\sim
N^{\phi-1}, 
\label{nsns}
\end{equation}
which should be valid for long chains ($N\gg1$).

In three dimensions, the precise value of this crossover
exponent still remains an open question, even after decades of intensive
research. For example, in the seminal work of Eisenriegler, Kremer, and Binder~\cite{eisen}
on scaling relations for the adsorption transition, the estimated value was
$\phi = 0.58(2)$. Meirovitch and Livne~\cite{meiro} have used a scanning
simulation method to obtain $\phi=0.530(7)$. On the other hand, Hegger and
Grassberger~\cite{hegger} suggested that this exponent might be superuniversal,
because they found $\phi= 0.496(5)$, which is close to the exact result in two
dimensions $\phi= 0.5$. Another result towards the superuniversal character of
this exponent has been reported by Metzger et al.~\cite{metz,metz2},
$\phi= 0.50(2)$. According to Descas et al.~\cite{descas}, the
determination of $\phi$ is strongly dependent on the 
estimation of the corresponding transition temperature $T_a$. In their work,
both values $\phi =0.5$ and $\phi = 0.59$ were considered acceptable, with the latter 
one being preferable. In a high-precision simulation using the
pruned-enriched Rosenbluth method (PERM), Grassberger~\cite{grass} 
obtained $\phi = 0.484(2)$. In agreement with this result,
Klushin et al.~\cite{binder} estimated $\phi = 0.483(3)$.
Conversely, Luo~\cite{luo} reported a larger value, $\phi = 0.54(1)$, 
while Taylor and Luettmer-Strathmann~\cite{taylor}, by means of Fisher partition 
function zeros, determined $\phi = 0.515(25)$. In a previous paper we have
obtained  $\phi = 0.492(4)$ \cite{pla} and, very recently, Bradly et al. settled at
$\phi =0.484(4)$ \cite{brad}.

From what has been discussed above, one can clearly notice that the estimates 
of the crossover exponent cover a broad range and they are 
strongly dependent on the precise value of the critical temperature $T_a$. 
Additionally, the
previous studies mostly consider good solvent conditions only, in which the
monomer-monomer interaction has been neglected. 
It is, therefore, interesting to see whether the
exponents (besides the crossover exponent $\phi$, there are critical exponents
for other thermodynamic quantities, which will be defined below)
vary or are universal as a function of the different solvent conditions,
together with the construction of the corresponding phase diagram. Put in this way, a 
major contribution of the present work is the discussion of the dependence of the 
critical behavior and the critical exponents under all possible solvent conditions, which
effectively extends the study to an entire class of hybrid polymer-
adsorbent-solvent systems instead of a single-case scenario of good solvent conditions as
has been done in the past.

In the present work, which is an extension of the recent results reported in 
reference \cite{pla}, we treat the critical properties of the A-D transition 
of long chain polymers described by a coarse-grained model of 
self-avoiding random walk in three dimensions (i.e. chains with excluded 
volume interactions)  with different solvent conditions by including an extra 
monomer-monomer interaction. 
In addition, we take advantage of the
similarity of this geometrical transition with those in magnetic systems, 
and perform a finite-size scaling analysis~\cite{luo,puli}.
However, corrections-to-scaling effects will be considered in order to 
take into account the finiteness of the polymers length. It will
be shown that the critical values are in fact in accord to some already
discussed in the literature for good solvents, not only for the 
estimates of the transition temperature $T_a$ but also for the 
corresponding crossover exponent as well.
For different values of the solvent conditions it is noted that the 
exponents vary and the presence of a multicritical point is expected along 
the phase transition line separating the adsorbed and desorbed phases.

The paper is organized as follows. In the next section the model is defined,
while the computer simulational background is briefly described in section
\ref{sim}. In section \ref{fss}, we present details of the finite-size-scaling 
analysis and the results are discussed in section \ref{result}. The last section
contains additional comments and concluding remarks.

\section{self-avoiding ramdom walk MODEL on the simple cubic lattice}
\label{mod}

A simple and useful coarse-grained lattice polymer model for adsorption can be
represented by an interacting self-avoiding random walk with additional
monomer-substrate interaction. A polymer chain of length $N$ is formed by $N$
identical monomers occupying sites on a simple cubic lattice. Adjacent monomers
in the polymer sequence have a fixed unitary bond length of one lattice unit. We
consider a grafted polymer with one end covalently, and permanently,  bound to
the surface (i.e. it cannot desorb).

Each pair of nearest-neighbor non-bonded monomers possesses an energy
$-\epsilon_m$. Thus, the key parameter for the energetic state of the
polymer itself is the number of monomer-monomer contacts,  $N_m$. The flat
homogeneous and impenetrable substrate is located in the $z=0$ plane, and the
polymer is restricted to $z>0$. All monomers lying in the $z=1$ plane are 
considered to be in contact with the substrate, and an energy $-\epsilon_s$ is
attributed to each one of these surface contacts. Hence, the energetic
contribution due to the interaction with the substrate is given by the number of
surface contacts of the polymer, $N_s$.

The total energy of the model can then be written as~\cite{bach05,bach14}
\begin{equation}
E_s(N_s, N_m) =-\epsilon_s N_s -\epsilon_m N_m= -\epsilon_s(N_s + s N_m), 
\label{en}
\end{equation}
where $s = \epsilon_m/\epsilon_s$ is the ratio of the respective
monomer-monomer and monomer-substrate energy scales.  
Actually, $s$ controls the solvent quality in such a way that 
larger $s$ values favor the formation of monomer-monomer contacts 
(poor solvent), whereas smaller values lead to a stronger binding to the substrate. 
For convenience, we set $\epsilon_s = 1$ meaning that all energies are 
measured in units of the monomer-substrate interaction energy scale.

\section{SIMULATIONAL BACKGROUND}
\label{sim}

We used the contact-density chain-growth algorithm~\cite{bach14,bach03,bach04}, 
where the  density of contacts $g(N_s, N_m)$ is directly obtained from the
simulation. This quantity represents the number of states 
for a given pair $N_s$ (number of contacts) and $N_m$ (monomer-monomer contacts). 
It does not depend on the temperature and the solvent parameter $s$. 
Thus, the temperature $T$ and the
solubility parameter $s$ are external parameters that can be set after the
simulation has finished.

We have simulated chains with lengths $N = 16,~ 32,~ 64,~ 128,~ 256,~ 400,~
\mbox{and}~ 503$ monomers. The total number of generated chains varied from
$3.0 \times 10^8$ for $N=16$ to $1.8 \times 10^9$ for $N=503$. Statistical
errors have been estimated by using the standard jackknife method~\cite{quen}.
The $s$ values we considered here varied from
$s=-10$ to $s=4$. It is noteworthy that other values of $s$ could be chosen
without performing any extra simulations. However, since we already covered a
significant region of the phase diagram, other values of solvent conditions
would not provide any new qualitative insights into the transition behavior.

It is interesting to note that all relevant energetic thermodynamic observables
can be obtained from the contact density $g(N_s, N_m)$. For
instance, for a given pair $N_s$ and $N_m$, one can define the restricted
partition function $Z_{T,s}^r (N_s, N_m)$ as
\begin{equation}
Z_{T,s}^r (N_s, N_m) = g(N_s, N_m) \exp[(N_s + s N_m)/k_B T ],
\end{equation}
from which the canonical partition function is obtained as
\begin{equation}
Z_{T,s} = \sum_{N_s,N_m}Z_{T,s}^r (N_s, N_m).
\end{equation}

Similarly, the mean value of any quantity $Q(N_s, N_m)$ can be computed from
\begin{equation}
\langle Q \rangle={{1}\over{Z_{T,s}}} \sum_{N_s,N_m} Q(N_s, N_m) g(N_s, N_m)
\exp\left[{{N_s + s N_m }\over{k_B T}} \right].
\end{equation}
In the simulations we set $k_B=1$.

It is then clear that entropy, free energy, the average number of surface
contacts $N_s$, the average number of monomer-monomer contacts $N_m$, heat
capacity, cumulants, etc. are examples of functions that are easily calculable
for any values of $T$ and $s$, as soon as $g(N_s, N_m)$ has been obtained from
the simulations.

Now, in order to get the critical properties of the present model, instead of
working with energetic quantities, such as the specific heat maximum~\cite{kraw}, 
which has been proven to have some pitfalls~\cite{janse}, or considering 
the scaling properties of the partition function~\cite{grass,binder}, 
we will resort here to the scaling properties of the order parameter, 
and its derivatives, along the same lines as done in reference~\cite{luo}. 
However, we will take into account corrections to scaling and
use convenient temperature derivatives of the order parameter 
$\langle n_s\rangle$, as well as
scaling properties of the A-D transition temperature and the fourth-order
cumulant of the order parameter.

So, from the simulations for each polymer length~$N$, we can compute the mean
value of the fraction of the number of monomer-substrate contacts, given by 
Eq.~(\ref{ns}), and related quantities like the fourth-order cumulant
\begin{equation}
 U_{4}(T)=1-\frac{\left<n_s^{4}\right>}{3\left<n_s^{2}\right>^{2}},
 \label{u4}
\end{equation}
and the temperature derivative
\begin{equation}
\Gamma= -\frac{d\ln\langle n_s\rangle}{dT}.
 \label{dlns}
\end{equation}

\section{Finite-size scaling (FSS) of the thermodynamic functions}
\label{fss}

\subsection{Magnetic systems}

According to the finite-size scaling (FSS) theory for second-order phase transitions, 
it is well known that the singular part of the magnetic Gibbs free energy $G(t,h,L)$ of a 
finite magnetic system, close to its critical temperature,  can be written as
\begin{equation}
G(t,h,L)=L^{-d}{\cal G}(L^{y_t}t,L^{y_h}h),
\label{g}
\end{equation}
where $t=|T-T_c|/T_c$, $T_c$ being the infinite lattice critical temperature, 
$h$ is the external 
field, $L$ is the linear system size, and $d$ is the dimension of the lattice. 
The exponents in Eq. (\ref{g}) are $y_t=1/\nu$ and $y_h=d-\beta/\nu$, where $\nu$ 
is the correlation length critical exponent and $\beta$ is the magnetization critical 
exponent. From the above relation, the scaling behavior of the magnetization, the
specific heat, and the susceptibility can be obtained in a straightforward way. 

From the free energy (\ref{g}) and, for simplicity, considering zero external field $h=0$, it 
can be shown that any of the  above specified quantities, generally designated by $P(t,L)$, 
scales as
\begin{equation}
P(t,L)=L^{\sigma/\nu}f_P(x),
\label{p}
\end{equation}
where $\sigma$ is the critical exponent of $P$, namely 
$P(t,L\rightarrow\infty)=P_0t^{-\sigma}$ (with $P_0$ a non-universal constant) and 
$f_P(x)$ is a FSS function of $x=L^{1/\nu}t$. For instance, if $P$ is 
the magnetization one would have $\sigma=-\beta$. Similarly, one has  $\sigma=\alpha$ and
$\sigma=\gamma$ for the specific heat and susceptibility, respectively.
 
The FSS ansatz given by Eqs. (\ref{g}) and (\ref{p}) is valid only for sufficiently large
systems and temperatures sufficiently close to the critical one. Corrections to scaling and 
finite-size scaling terms should appear for smaller systems and temperatures away from
$T_c$, mainly due to irrelevant scaling fields and non-linear scaling fields. 
In general, such corrections can be implemented by generalizing (\ref{p}) as
\begin{equation}
P(t,L)=L^{\sigma/\nu}f_P(x)\left[1+A_{P}(x)L^{-\omega}\right],
\label{pc}
\end{equation}
where $A_P(x)$ is another FSS function of $x=L^{1/\nu}t$ and $\omega$ is the 
corresponding leading order correction-to-scaling exponent.

\subsection{Adsorbed polymer chain}

For the present coarse grained lattice polymer model for adsorption the natural size of 
the self-avoiding random walk chain is its length $N$. In analogy with magnetic systems,
as given by Eq. (\ref{g}), the polymer free energy should scale with temperature as
\begin{equation}
G(t,N)=N^{-1}{\cal G}(N^{1/\delta}t),
\label{gp}
\end{equation}
where we have now $y_t=1/\delta$ instead of $y_t=1/\nu$ in order to avoid confusion with
Flory's $\nu$ exponent widely used in polymer science.

The general scaling relation for the order parameter (\ref{nsns}) reads
\begin{equation}
\langle n_s\rangle =N^{\phi-1}f_{n_s}(x)\left[1+A_{n_s}(x)N^{-\omega}\right],
\label{nscal}
\end{equation}
where the above equation is a generalization of Eq.~(\ref{nsns}) by taking into
account effects of corrections to scaling due to the finiteness of the polymer
length and now $x=N^{1/\delta}(T-T_a)$, $T_a$ being the adsorption critical temperature. 

The corresponding fourth-order cumulant of the order parameter 
$U_4$ given by Eq.~(\ref{u4}) (also known as Binder cumulant) should be 
independent of the chain length $N$ (for very long chains)~\cite{bincum}, 
and the maximum value of the quantity given by Eq.~(\ref{dlns}) is supposed to scale as
\begin{equation}
{\Gamma}_{max}=- \left[{\frac{d\ln\langle
n_s\rangle}{dT}}\right]_{max}=N^{1/\delta}f_d(x)\left[1+A_d(x)N^{-\omega}\right].
\label{dscal}
\end{equation}

Accordingly, for the critical temperature one has the following scaling law,
also based on continuous transitions in magnetic models,
\begin{equation}
T_N= T_a + N^{-1/\delta}f_{T}(x)\left[1+A_{T}(x)N^{-\omega}\right].
\label{tscal}
\end{equation}

The exponent $\delta$ defined above is in fact the equivalent of the thermal critical exponent
of the correlation length $\nu$ in ordinary magnetic continuous phase
transitions. The functions $f_i(x)$, with $i=n_s,~d,~T$, are FSS functions and $A_i(x)$ 
are non-universal functions (see, for instance, reference~\cite{landaupa}).

Thus, the procedure we have to follow now is quite simple. From the simulations
of each polymer size, we determine the exponent $1/\delta$ by using 
Eq.~(\ref{dscal}), which depends only on the maximum value of the derivative given
by Eq.~(\ref{dlns}). In this case, we consider $f_d(x)$ and $A_d(x)$ as normal
constants [we expect them not to vary much since the maximum positions
should occur at temperatures close to the critical one so that one has $f_d(x=0)$ and 
$A_d(x=0)$]. With this exponent in
hands, the critical temperature $T_a$ is obtained from Eq.~(\ref{tscal}) and,
having the critical temperature, we are able to get the crossover exponent
$\phi$ by using Eq.~(\ref{nscal}), where in this case we can also consider $x=0$.

\section {RESULTS}
\label{result}

As a matter of comparison, and a test for the performance of the present
approach, let us first discuss the corresponding results for the good solvent
condition case $s=0$, where we have previous results originating from different
methods. 

\subsection{Good solvent condition s=0}

\begin{figure}[!t]
   \centering
   \includegraphics[clip,angle=0,width=8.5cm]{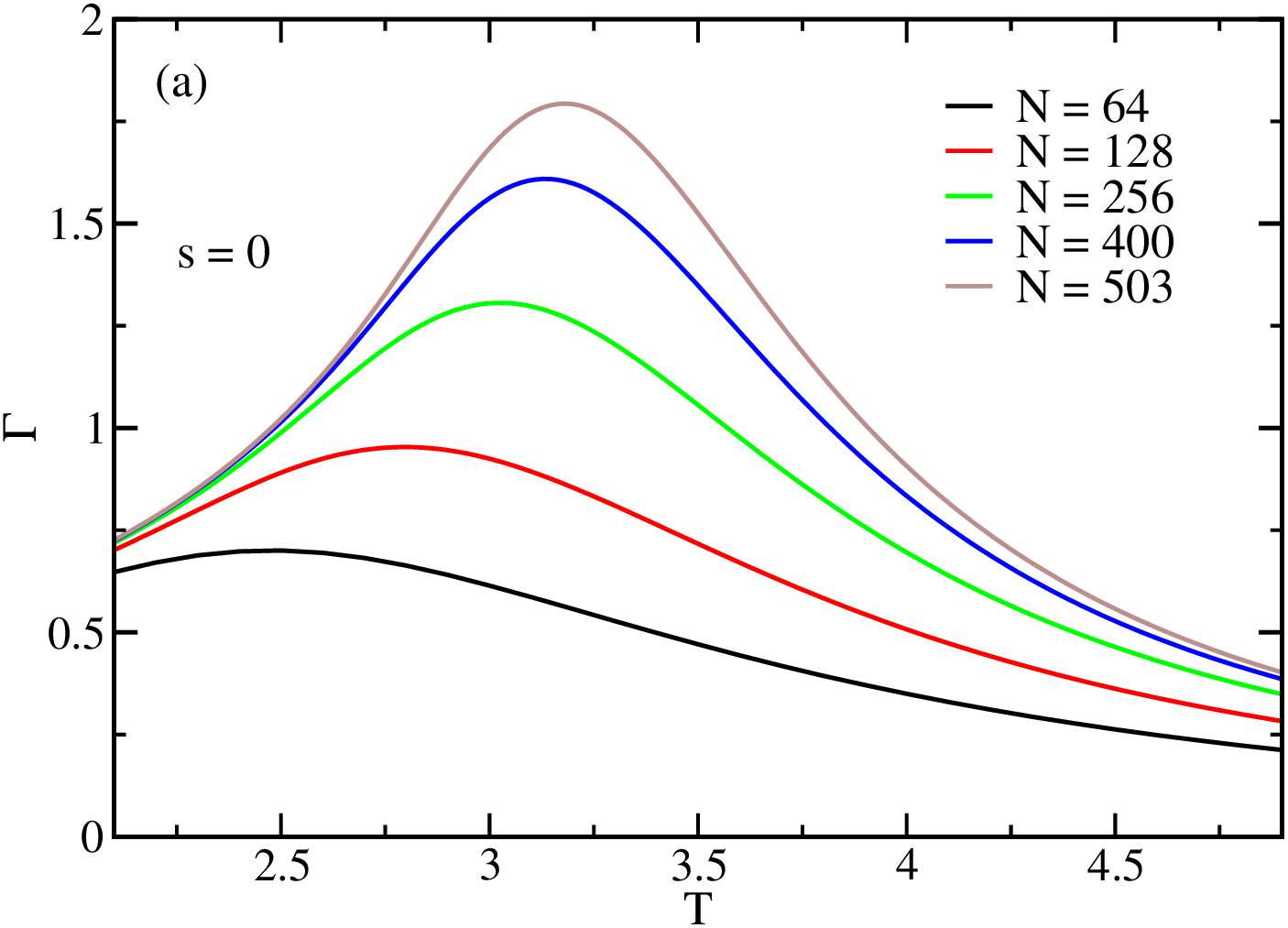}
\end{figure}
\begin{figure}[!t]
   \centering
   \includegraphics[clip,angle=0,width=8.5cm]{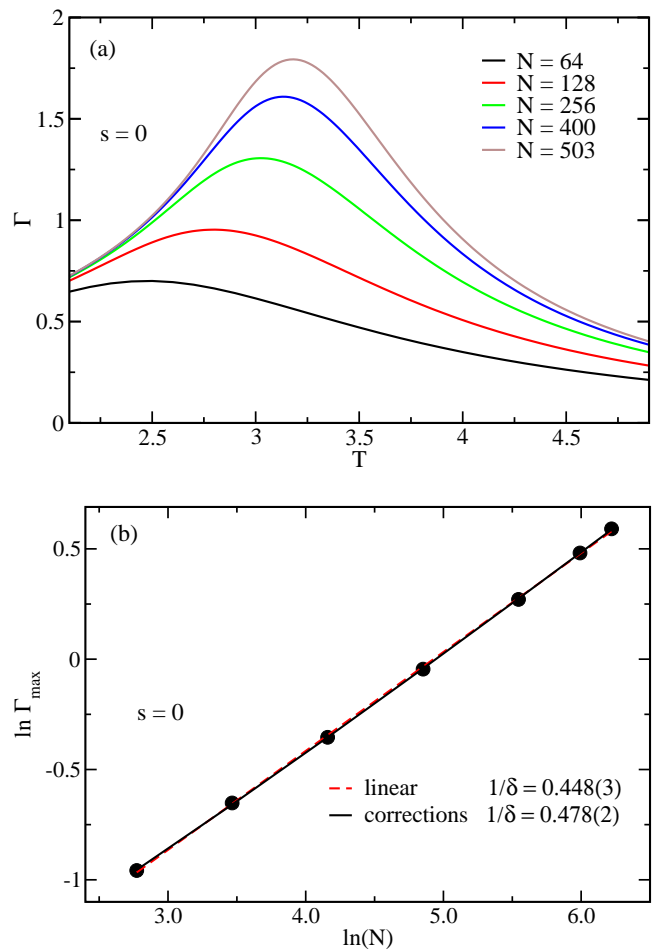}
\caption{ (color online) Results for $s=0$. (a) $\Gamma$, as defined in 
Eq. (\ref{dlns}), as a function of temperature, for different polymer sizes. Smaller sizes
have been omitted for clarity.
(b) Logarithm of the maximum value of $\Gamma$ obtained in (a) as a function of 
the logarithm of the polymer
length $N$ for different chain sizes. The dots correspond to the
simulation results and the lines are the best fit according to Eq.~(\ref{dscal}), 
without corrections to scaling (linear, taking $A_d(x=0)=0$) and
with corrections to scaling ($A_d(x=0)\ne 0$). The error bars are smaller than the symbol sizes.}
\label{s0nu}
\end{figure}

Fig. \ref{s0nu} shows results obtained by using Eqs. (\ref{dlns}) and 
(\ref{dscal}). In Fig. 
\ref{s0nu}(a) we have plotted $\Gamma$ as a function of temperature for several polymer
lengths. The corresponding logarithm of its maximum value as a function of the 
logarithm of the polymer length $N$ for different chain sizes is shown in Fig.
\ref{s0nu}(b). From the linear fit one gets $1/\delta=0.448(3)$,
while taking into account corrections to scaling one gets $1/\delta=0.478(2)$.
Although the corresponding data are rather close, as can be
seen in Fig. \ref{s0nu}(b), the value of the $1/\delta$ exponent is sensitive  when one
considers correction to scaling. This value should be compared to the estimate
$1/\delta=0.56$ from reference~\cite{luo} and $1/\delta=0.485(6)$ from reference \cite{brad}, 
where the latter ones have been obtained from different procedures. 

\begin{figure}[htb]
 \includegraphics[angle=0,width=8.5cm]{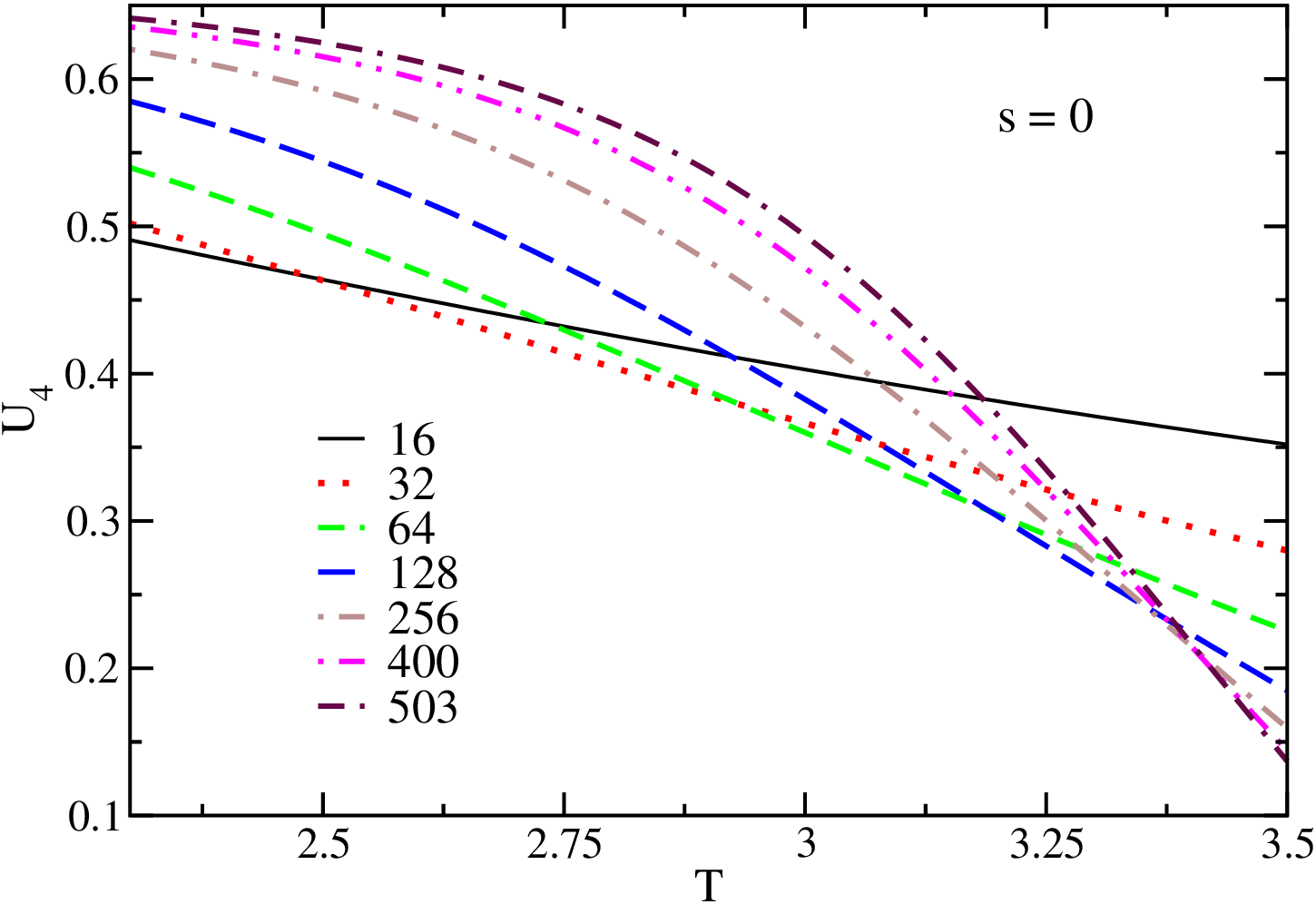}
\caption{(color online) {Fourth-order Binder cumulant $U_4$ as a function of
the temperature $T$ for different chain sizes for $s=0$.}} 
\label{cums0}
\end{figure}

\begin{figure}[htb]
\includegraphics[angle=0,width=8.5cm]{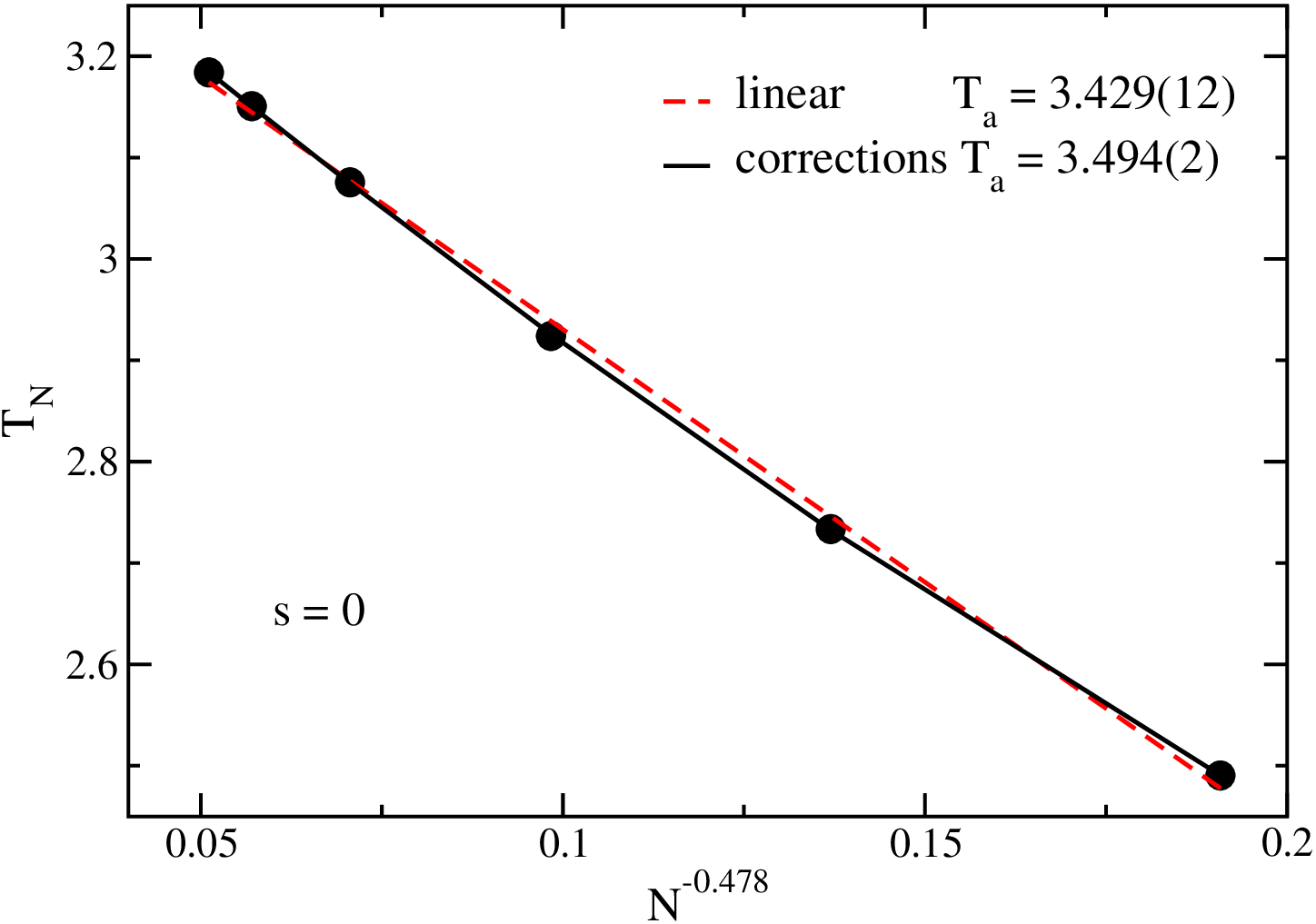}
\caption{(color online) { Transition temperature $T_N$ as a function of
$N^{-1/\delta}$ for $s=0$. The dots correspond to the crossings of the
fourth-order Binder cumulant for chain lengths $N\ge32$ with the result for
$N=16$, as shown in Fig.~2.
The lines are the best fit according to Eq.~(\ref{tscal}), 
without corrections to scaling (linear, taking $A_T(x=0)=0$) and
with corrections to scaling ($A_T(x=0)\ne 0$).}}
 \label{s0ta}
\end{figure}

The fourth-order Binder cumulant, as a function of the temperature, is shown in
Fig. \ref{cums0}.
 One can clearly see that there is a systematic crossing of the
larger chains with relation to the smallest one ($N=16$).  Taking these crossings
as $T_N$, for $N\ge32$, we can plot them as a function of $N^{-1/\delta}$,
where $1/\delta$ has been computed from the data of Fig. \ref{s0nu}. 
The corresponding results are depicted in Fig. \ref{s0ta}. 
We note that corrections to scaling are more important in this case and the value 
$T_a=3.494(2)$, so calculated, is comparable to the values  
$T_a=3.500(1)$ obtained by Klushin et al.~\cite{binder}, $T_a=3.44(2)$ by Luo \cite{luo}, 
and the most recent estimate $T_a=3.520(6) $ obtained by 
Bradly et al. \cite{brad}. All these estimates were obtained by different approaches. 

\begin{figure}[htb]
\includegraphics[angle=0,width=8.5cm]{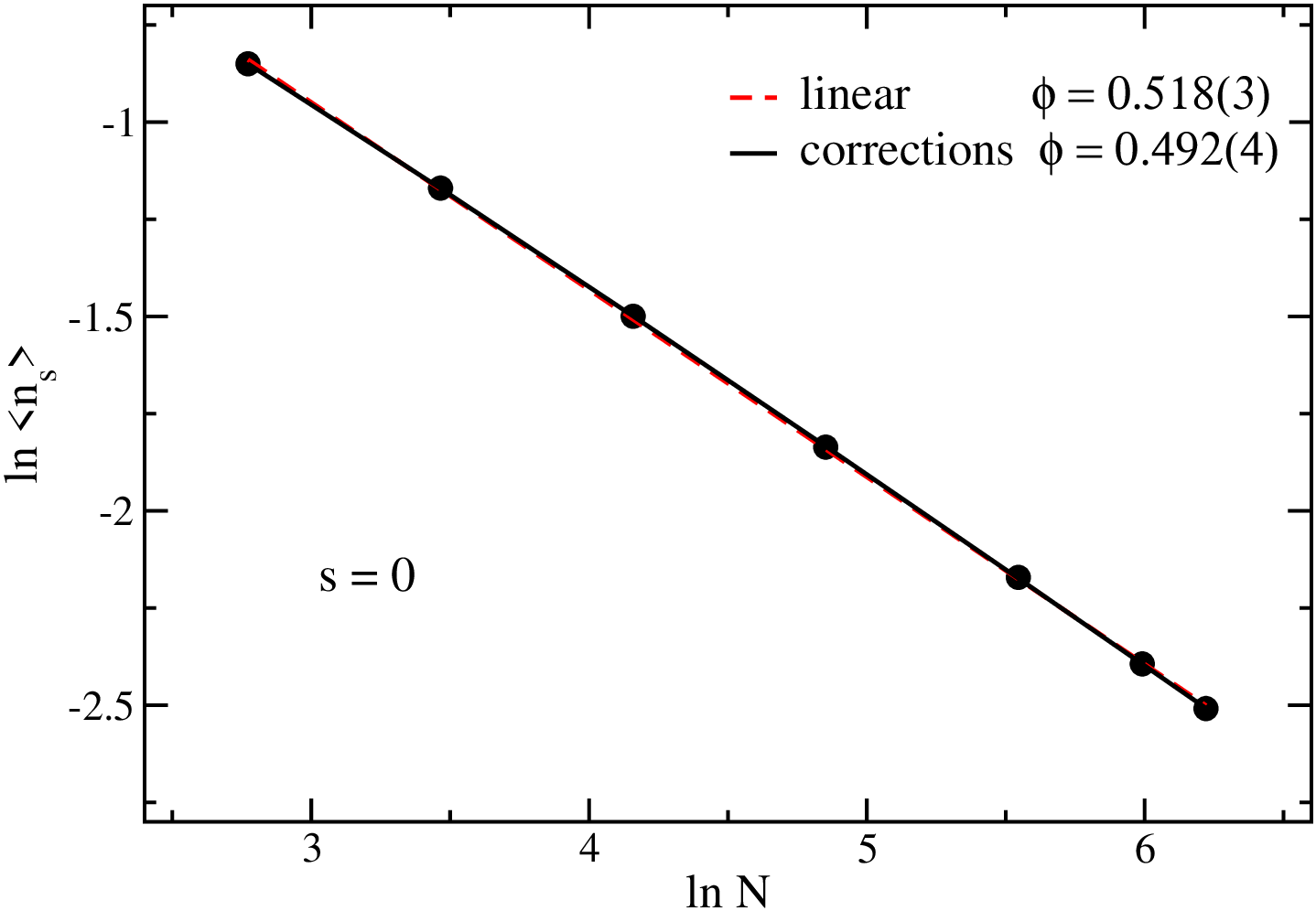}
\caption{ (color online){Logarithm of the order parameter $\langle n_s\rangle$ as a function
of the logarithm of the polymer length $N$ for different chain sizes, for $s=0$,
at the transition temperature $T_a$. The dots correspond to the
simulation results and the lines are the best fit according to Eq.~(\ref{nscal}), 
without corrections to scaling (linear, taking $A_{n_s}(x=0)=0$)
and with corrections to scaling ($A_{n_s}(x=0)\ne 0$). }}
\label{s0phi}
\end{figure}

Once the critical temperature has been calculated, one can now utilize the
scaling relation~(\ref{nscal}) to get the crossover exponent. The results are
shown in Fig. \ref{s0phi}. 
Although not visible in the scale of the figure,
the corrections to scaling are important in this case as well, and the computed value
$\phi=0.492(4)$ is also in agreement with  the result quoted in 
reference \cite{brad}, $\phi=0.485(6)$, and quite close to the value $\phi=0.483(2)$
given in reference~\cite{binder} (these results are smaller than the estimate $0.56$ obtained 
by Luo \cite{luo}).

From the above results, we can see that the present approach can furnish quite
good results for the special case $s=0$, when compared to the values previously
obtained from other procedures. The corresponding values of the critical
behavior are displayed in Table~\ref{tab}, together with those from 
references~ \cite{binder}, \cite{luo}, and \cite{brad} for a comparison. 
However, our 
approach has the advantage of being easily extended to get the critical behavior
for other values of the solvent condition parameter $s$ without any extra
simulation.

\begin{table}
\caption{Adsorption critical temperature $T_a$ and crossover exponents $\phi$ 
and $1/\delta$ for some selected values of the solvent conditions $s$. For comparison,
some values for $s=0$ coming from different methods are also shown. The last row gives
the estimate of the correction-to-scaling exponent.}
\begin{tabular}{c c c c }
\hline \hline
$s$  &   $T_a$   &   $\phi$   &1/$\delta$ \\
\hline
-10 & 3.31(1)~  & 0.469(5) & 0.44(1)~~ \\
\hline
~~-5~~ &~~3.303(4)~~~&~~ 0.473(3)~~~ &~ 0.448(1)~~~ \\ 
\hline
-2 & 3.358(4)  & 0.478(3) & 0.450(8) \\
\hline
-1 & 3.407(3)  & 0.482(3) & 0.453(8) \\
\hline\hline
 0 & 3.494(2)  & 0.492(4) & 0.478(2) \\ 
 \hline
~~~~0\cite{binder} & 3.500(1)  & 0.483(2) &-- \\ 
 \hline
~~~~0\cite{luo}     & 3.44(2)~~    & 0.54(1)~~    & 0.56~~~~~ \\ 
 \hline
~~~~0\cite{brad}     &  3.520(6)   &  0.484(4)   & =$\phi$~~~~~~ \\
\hline\hline
 1 & 3.788(9)  & 0.524(4) & 0.59(3)~~\\ 
 \hline
 ~~1.5 & 4.60(2)~~  & 0.353(4) & 0.52(1)~ \\
\hline
 2 & 5.74(2)~~  & 0.228(2) & 0.39(1)~~\\ 
 \hline
~~2.5 & 6.87(7)~~  & 0.20(3)~~ & 0.29(1)~ \\
\hline
3 & 7.9(1)~~~~  & 0.205(2) & 0.232(7) \\
\hline
 &   & $\omega=0.5(1)$ &  \\
\hline
\hline
\end{tabular}
\label{tab}
\end{table}

\subsection{Different solvent conditions $s\ne 0$} 

\begin{figure}[!t]
   \centering
   \includegraphics[clip,angle=0,width=8.5cm]{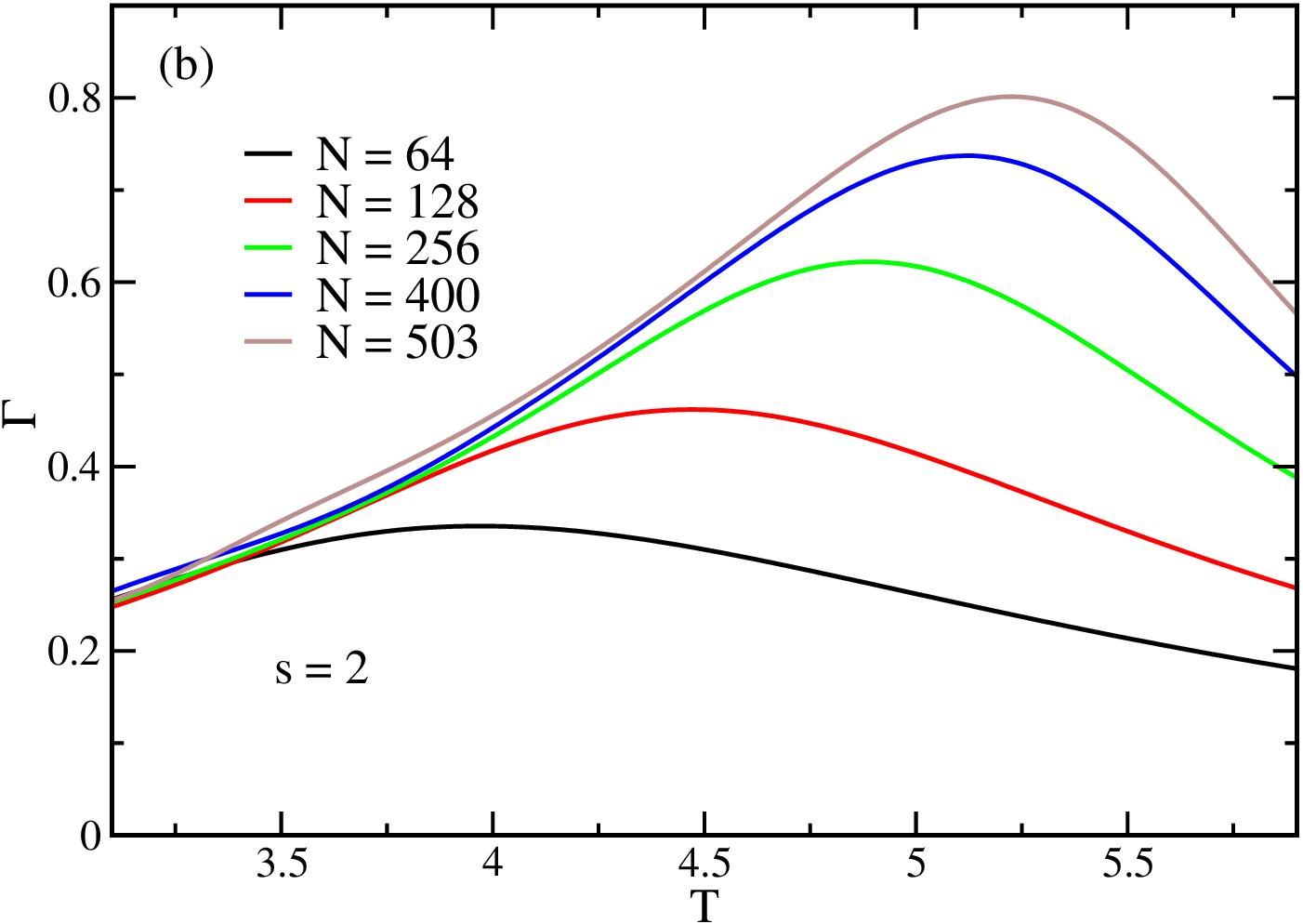}
\caption{ (color online) $\Gamma$, as defined in 
Eq. (\ref{dlns}), as a function of temperature, for different polymer sizes. 
In (a) we have $s=-2$ and in (b) $s=2$. In both cases, smaller sizes have 
been omitted for clarity.}
\label{gas-22}
\end{figure}

\begin{figure}[htb]
 \includegraphics[angle=0,width=8.5cm]{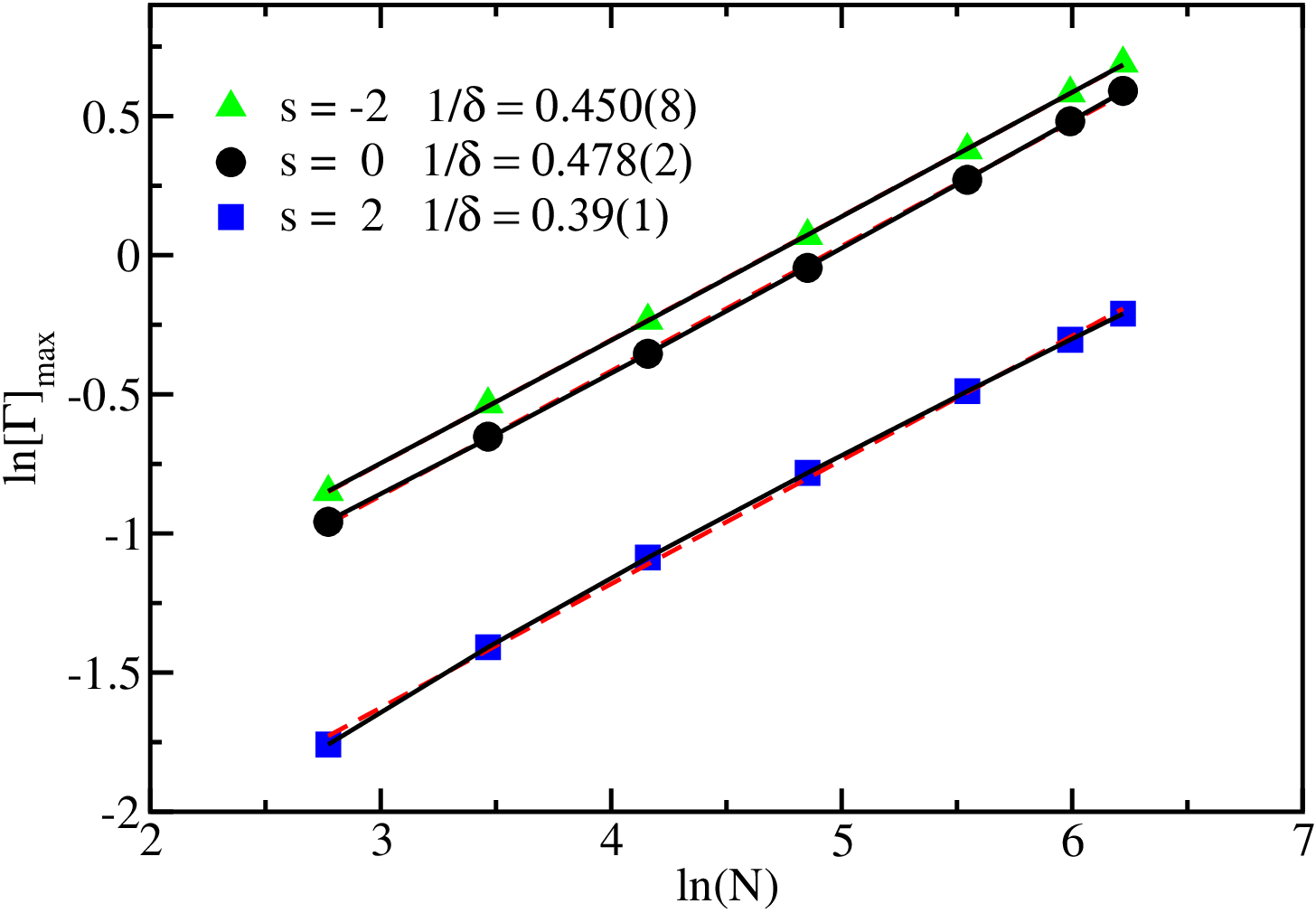}
\caption{(color online) {Logarithm of the maximum value of $\Gamma$ obtained from 
the data of Fig. \ref{gas-22}, as a function of the logarithm of the polymer
length $N$, for different chain sizes. The dots correspond to the
simulation results and the lines are the best fit according to Eq.~(\ref{dscal}). 
Dashed lines without corrections to scaling (linear, taking $A_d(x=0)=0$) and full lines
with corrections to scaling ($A_d(x=0)\ne 0$). Only the values coming from correction 
to scaling are listed in the figure. The error bars from the simulations are smaller 
than the symbol sizes.}} 
\label{nus-202}
\end{figure}

The behavior of the thermodynamic quantities for other values of the solvent
conditions are qualitatively similar to those presented in Figs. \ref{s0nu}-\ref{s0phi}. 
For example, the behavior of $\Gamma$
as a function of temperature is shown for different polymer chains for $s=-2$ in Fig. 
\ref{gas-22}(a) and for $s=2$ in  Fig. \ref{gas-22}(b).  The logarithm of the 
maximum value of $\Gamma$, as a function of the logarithm of the polymer length $N$, 
for the different chain sizes, is shown in Fig. \ref{nus-202} for both values of 
solvent conditions, together with the previous data of good solvent for comparison.
The values of the exponent $1/\delta$ include corrections to scaling.

\begin{figure}[!t]
   \centering
   \includegraphics[clip,angle=0,width=8.5cm]{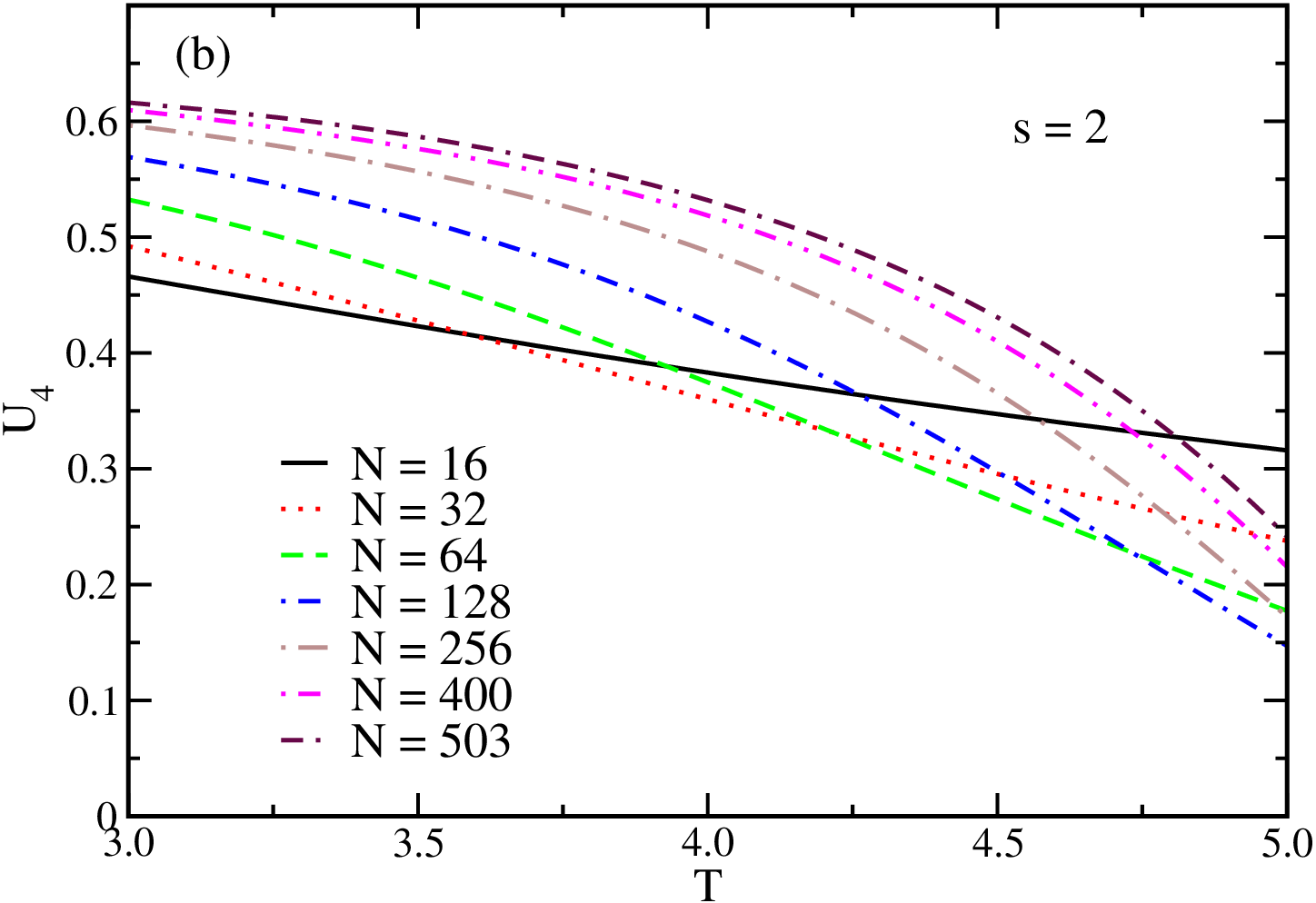}
\caption{ (color online) Fourth-order Binder cumulant $U_4$ as a function of
the temperature $T$ for different chain sizes for (a) $s=-2$ and (b) $s=2$.}
\label{binds-22}
\end{figure}

\begin{figure}[htb]
\includegraphics[angle=0,width=8.5cm]{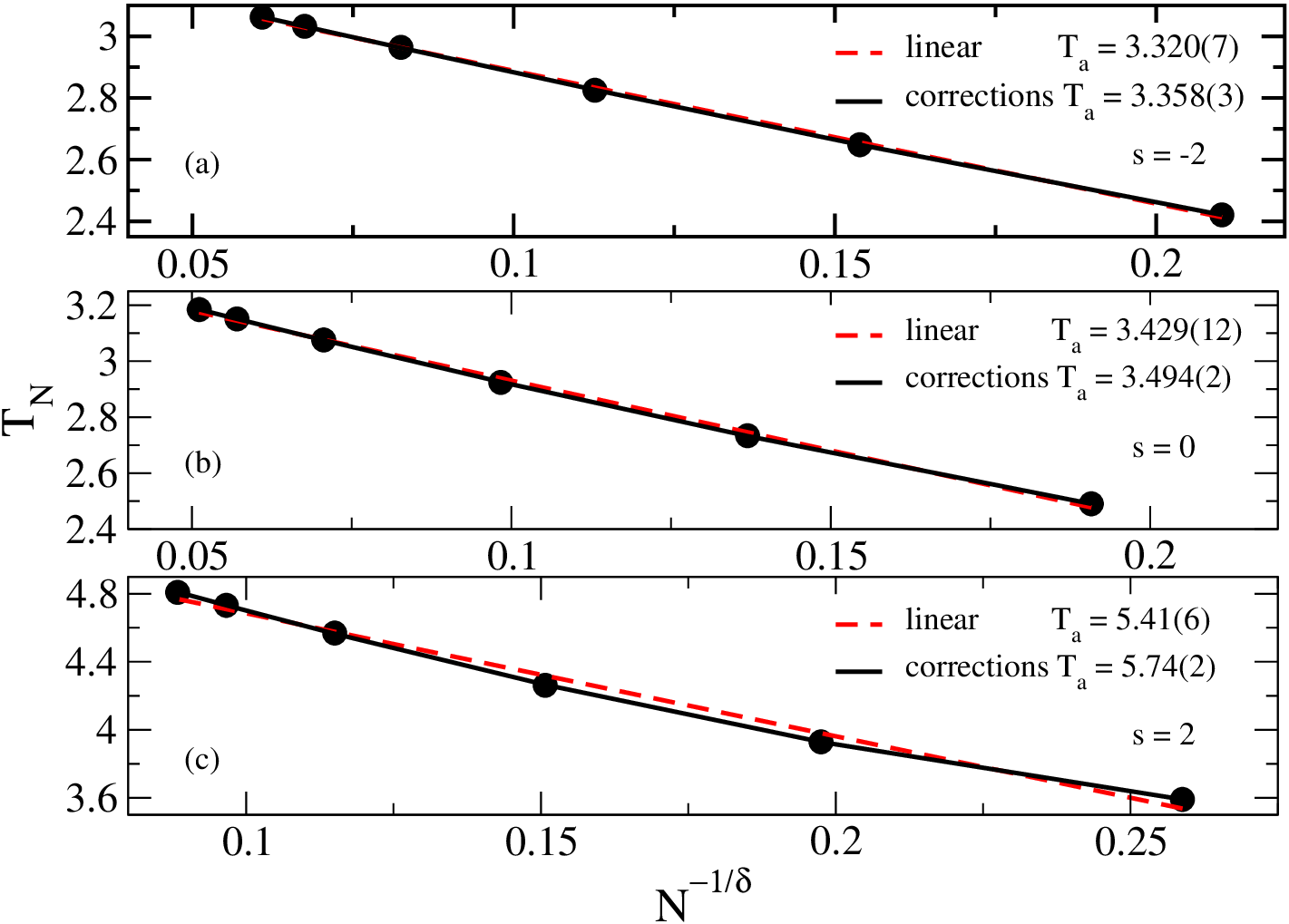}
\caption{(color online) { Transition temperature $T_N$ as a function of
$N^{-1/\delta}$ for $s=-2$ (a), $s=0$ (b), and $s=2$ (c). The values of the
exponents $1/\delta$, for each solvent condition, come from the data of Fig. \ref{nus-202}.
The dots correspond to the crossings of the
fourth-order Binder cumulant for chain lengths $N\ge32$ with the result for
$N=16$, as shown in Figs. \ref{cums0} and \ref{binds-22}. 
The lines are the best fit according to Eq.~(\ref{tscal}). Dashed lines 
without corrections to scaling (linear, taking $A_T(x=0)=0$) and full lines
with corrections to scaling ($A_T(x=0)\ne 0$).}}
 \label{tcs-202}
\end{figure}

Fig. \ref{binds-22} depicts the behavior of the fourth-order Binder cumulant for $s=-2$
[Figure \ref{binds-22}(a)] and $s=2$ [Figure \ref{binds-22}(b)]. From these curves
the crossings of the cumulants with the smaller chain ($N=16$) can be determined in order
to compute $T_N$ for $N\ge 32$ under each solvent condition. The corresponding results are
presented in Fig. \ref{tcs-202} together with the respective fits and extrapolated
adsorbed critical temperature $T_a$. The previously obtained value of the adsorbed transition
temperature for $s=0$ is also included in Fig. \ref{tcs-202} for comparison. We note 
that the scale of the curves for $s=-2$ and $s=0$ are the same, showing that there is 
not a sensitive variation in $T_a$ as for $s=2$.
\begin{figure}[!t]
   \centering
   \includegraphics[clip,angle=0,width=8.5cm]{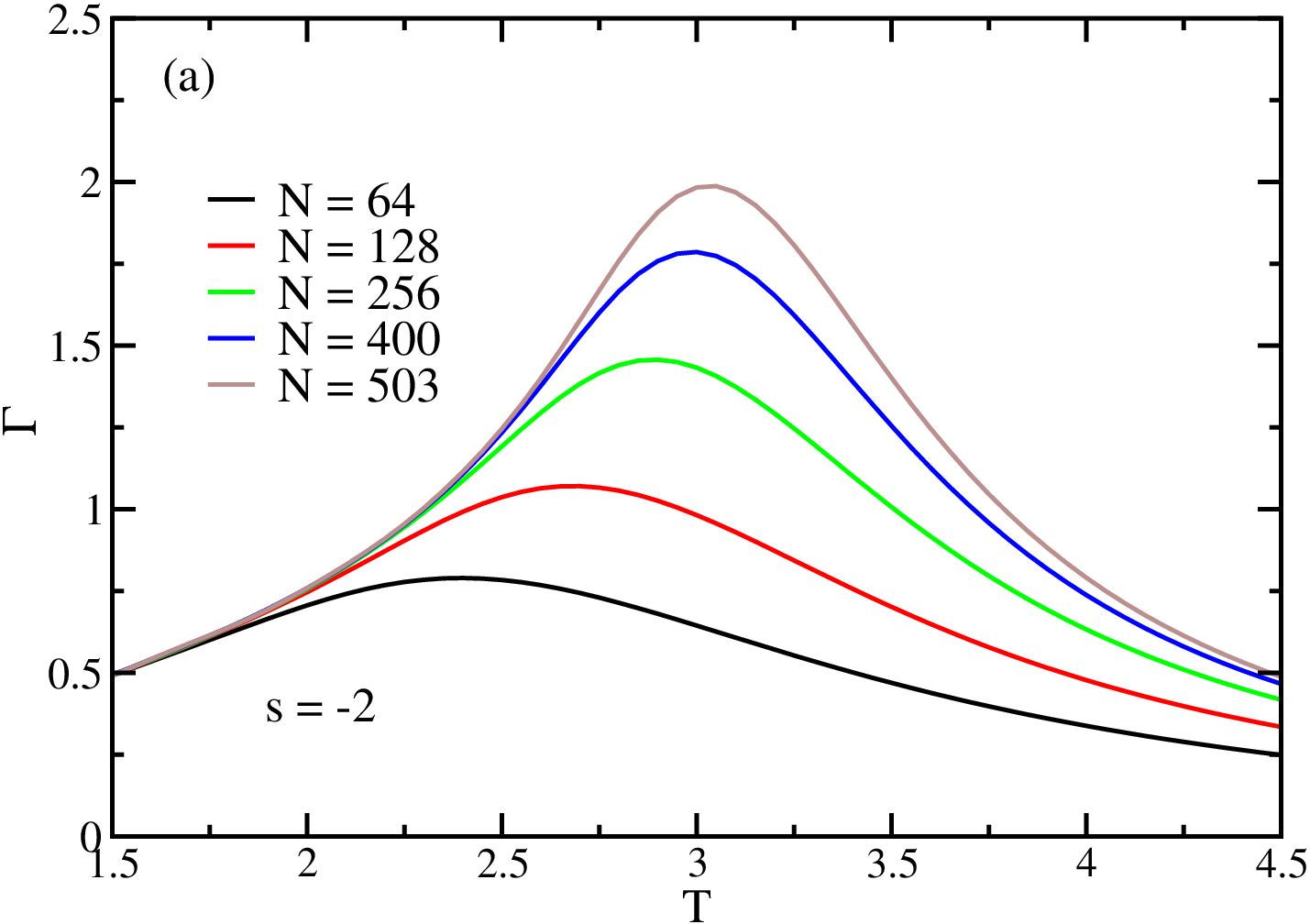}
\end{figure}

\begin{figure}[!t]
   \includegraphics[clip,angle=0,width=8.5cm]{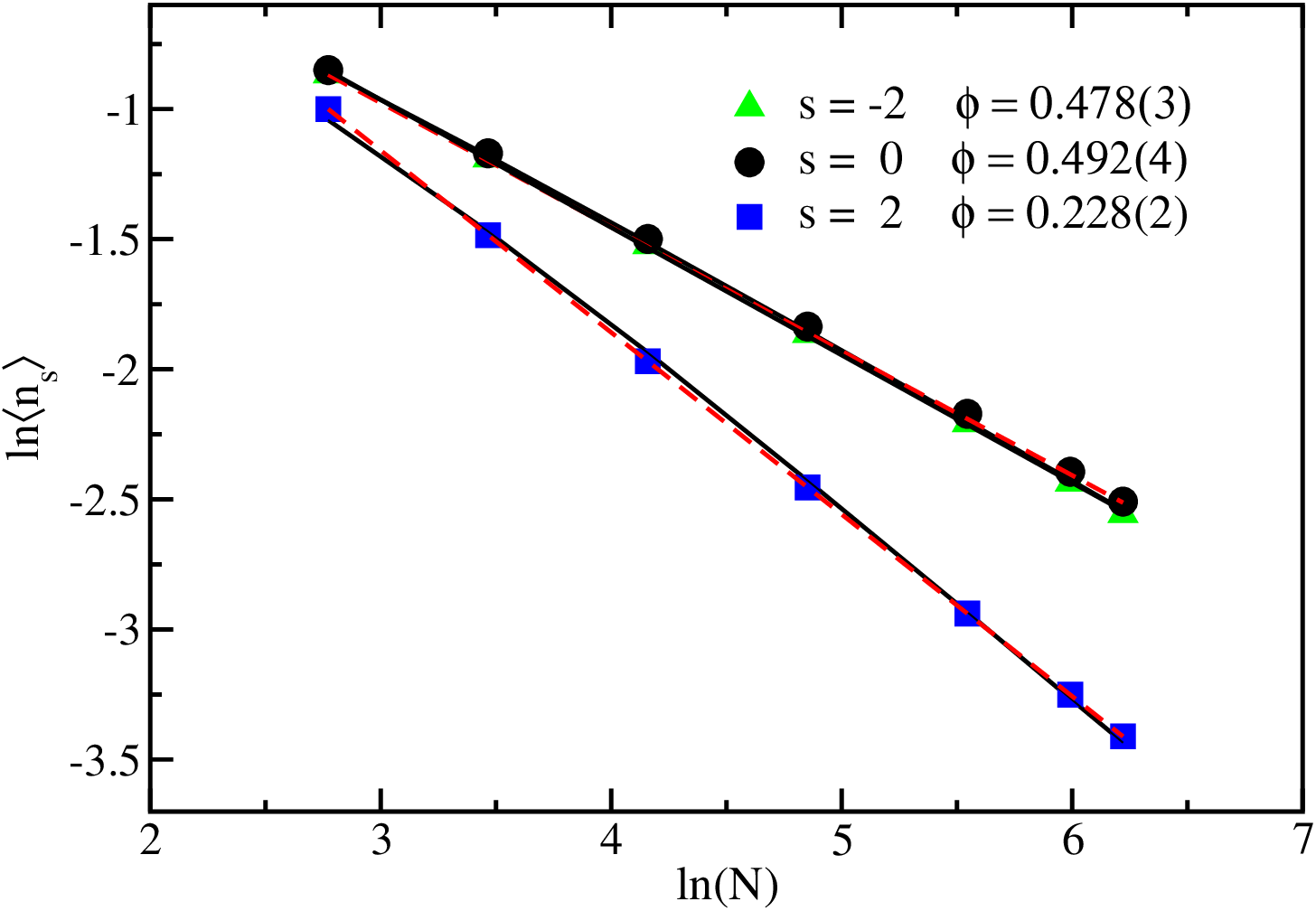}
\caption{ (color online) 
 Logarithm of the order parameter $\langle n_s\rangle$ as a function of the logarithm of 
the polymer length N, for different chain sizes, at the transition temperature $T_a$, obtained
from Fig. \ref{tcs-202}. The dots correspond to the
simulation results and the lines are the best fit according to Eq.~(\ref{tscal}). 
Dashed lines without corrections to scaling (linear, taking $A_d(x=0)=0$) and full lines
with corrections to scaling ($A_d(x=0)\ne 0$). Only the values coming from correction 
to scaling are listed in the figure. The error bars from the simulations are smaller 
than the symbol sizes.}
\label{phis-202}
\end{figure}

With the values of $T_a$ we can proceed and compute the crossover exponent $\phi$. The
results are shown in Fig. \ref{phis-202}. In all the fits one can notice that 
corrections to scaling are indeed important in obtaining the critical behavior of the
polymer. For easier comparison, all results are compiled 
in Table \ref{tab}, together with some additional selected values for different solvent conditions 
$s$.

\subsection{Discussion of the phase diagram and critical behavior}

\begin{figure}[htb]
  \includegraphics[angle=0,width=8.5cm]{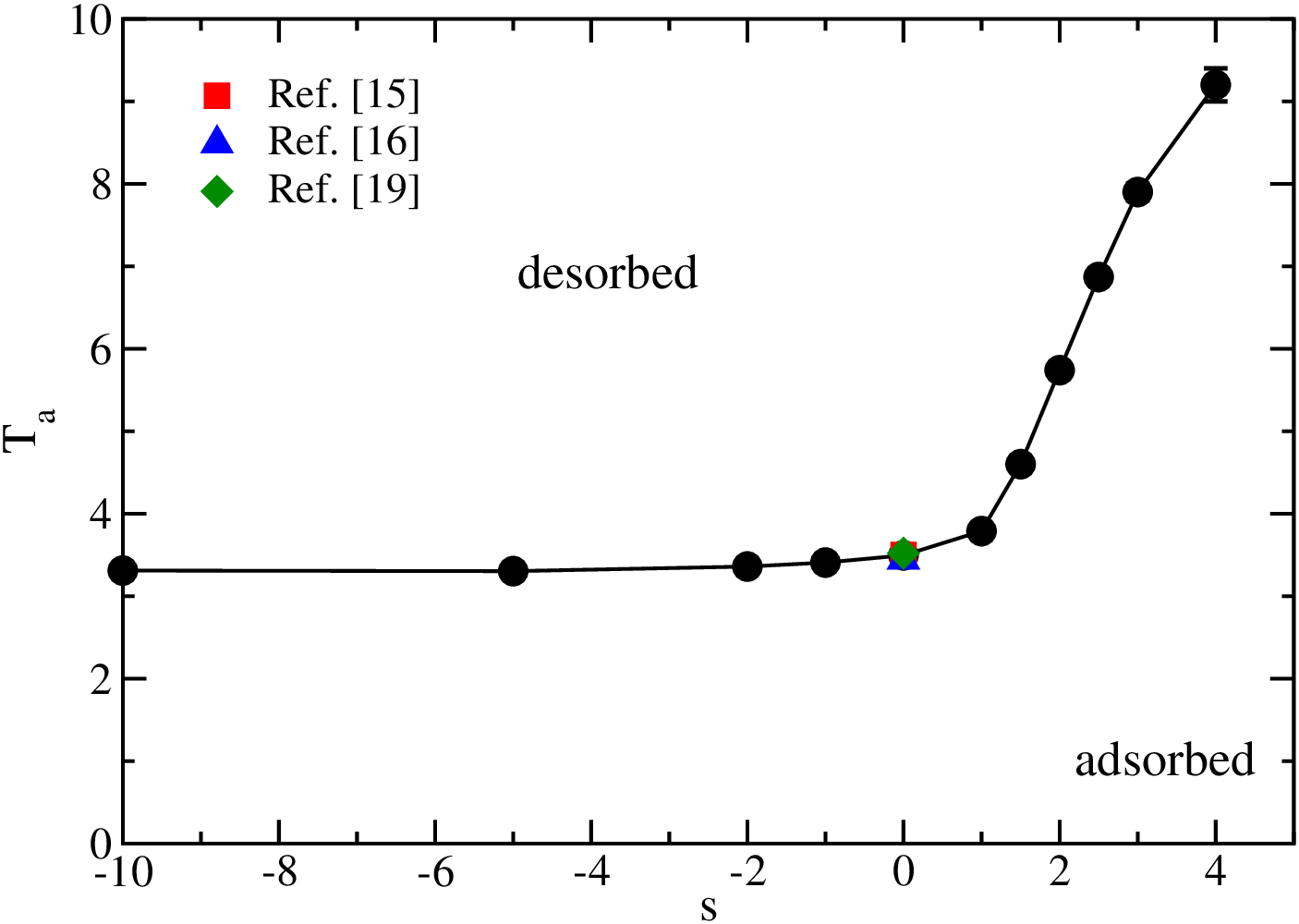}
  \caption{Critical temperature $T_a$ as a function of $s$ for the
adsorbed-desorbed transition.  Results for $s=0$ from references~\cite{binder,luo,brad} 
are also shown for comparison. The line is just a guide to the eyes.}
  \label{Tcs}
\end{figure}

Although the scaling behavior of the thermodynamic quantities for general solvent conditions
($s\ne0$) is qualitatively similar to that for good solvent ($s = 0$) the character of the adsorption 
transition changes considerably. This is clearly seen in the phase
diagram shown in Fig. \ref{Tcs}. Results for $s = 0$ from Refs.
\cite{binder}, \cite{luo}, and \cite{brad}, also included in this figure, fit very well into the
extended picture of polymer adsorption presented here. For  poor solvent ($s > 0$), 
desorbed and adsorbed polymer conformations are much more compact. The self-interacting
polymer undergoes a collapse and an additional freezing transition, and both transitions 
compete with the adsorption transition, depending on the solvent conditions. From the
estimates for transition temperatures and critical exponents,
we find that the specific parametrization of the critical behavior
depends in fact on the solvent quality. As Table \ref{tab} shows, the values of
the exponents obtained for $s\ne 0$ are significantly
different. Obviously, the solvent quality has a noticeable
quantitative influence on the adsorption behavior.
\begin{figure}[!t]
   \centering
   \includegraphics[clip,angle=0,width=8.5cm]{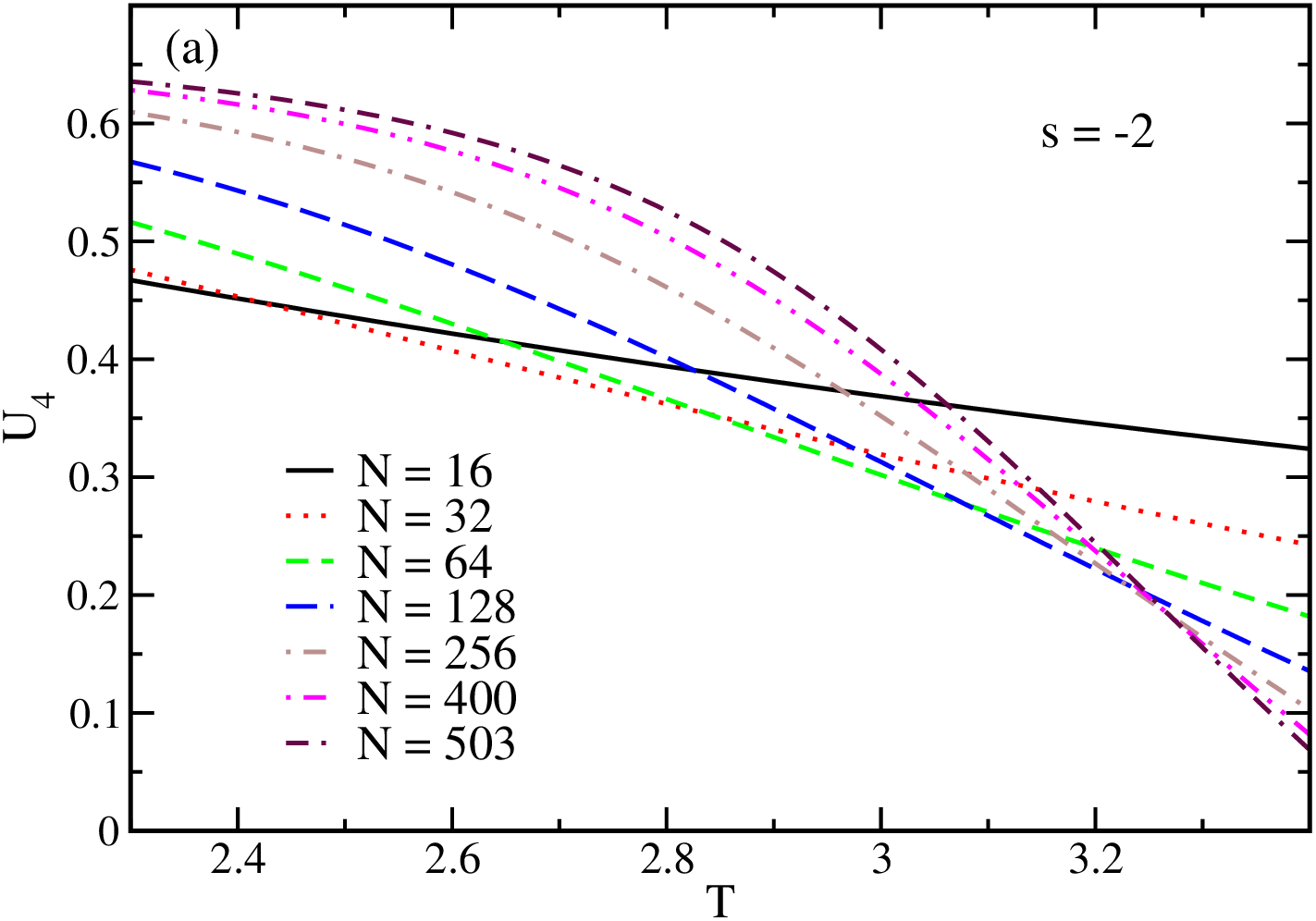}
\end{figure}

On the other hand, if $s$ is negative, the monomer-monomer interaction is
repulsive, and the polymer avoids forming nearest-neighbor
contacts. This mimics the effect of a good solvent. In the limit
$s\rightarrow-\infty$, the system is represented by what we may call a
“{\it super-self-avoiding walk}” (SSAW) model, where the contacts
between nearest neighbors are forbidden. This effectively
increases the excluded volume and the corresponding adsorption temperature
of the system is expected to be smaller than for $s = 0$.
To our knowledge, this case has not yet been studied and
there are no values to compare with. However, as our results
suggest, the corresponding critical adsorption temperature of
this intrinsically non-energetic SSAW should be $T_a \lesssim 3.31$.

Increasing the value of $s$  effectively
increases the conformational entropy at a given energy in
the phase of adsorbed conformations more than in the desorbed
phase. As a consequence, the slope of the microcanonical entropy
(or the density of states) becomes smaller near the transition
point, which, in turn, results in a larger adsorption temperature.
The phase diagram plotted in Fig. \ref{Tcs} shows exactly this behavior
for the adsorption temperature. 

\begin{figure}[htb]
   \includegraphics[angle=0,width=8.5cm]{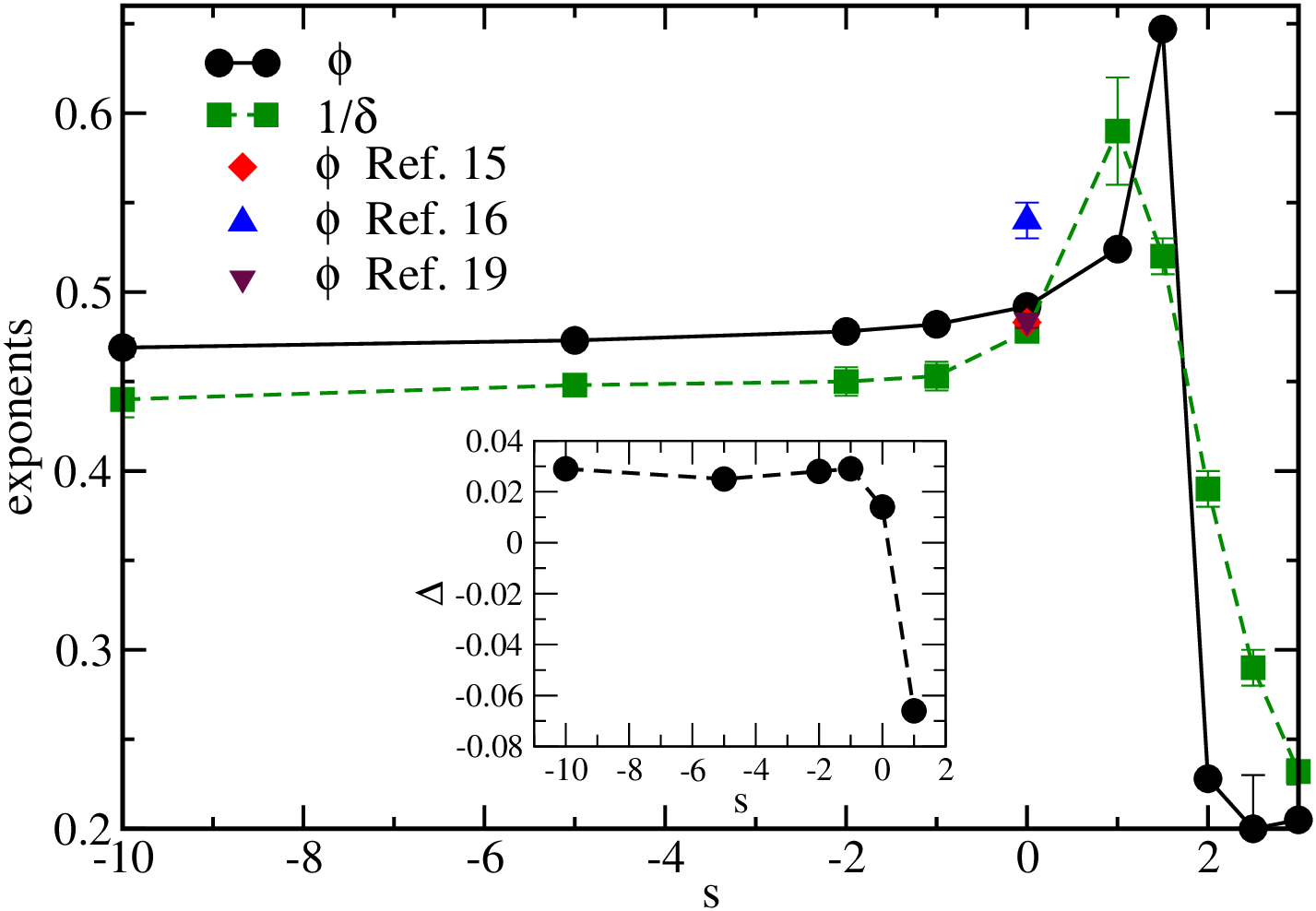}
\caption{(color online) {Critical exponents $\phi$ and $1/\delta$ as a function of the
solvent parameter $s$. Results for $s=0$ from references~\cite{binder,luo,brad} 
are also shown for comparison. The lines are guides to the eyes only. The inset
shows the difference between the exponets $\Delta=\phi-1/\delta$ as a function of $s$.}}
\label{phinu}
\end{figure}

For all $s$ values, the adsorption transition is of second-order.
Figure \ref{phinu} depicts
the behavior of the exponents $\phi$ and $1/\delta$ if the solvent quality $s$ is 
changed. We find that their values vary along the second-order transition
line, meaning that this transition seems to be non-universal.
Moreover, they cross each other close to $s=0$ and both exponents exhibit a peak near 
$s \sim 1.5$. These peaks can be an indication of the presence of a multicritical point
in this region \cite{bell,vrb,raj,owc}. In fact, various additional crossovers
between different adsorbed phases in the high-$s$ regime are
expected. Analyses for a finite system \cite{puli2} show a complex
structure of adsorbed compact phases in this regime, but
simulations of sufficiently large systems which would allow
for a thorough finite-size scaling analysis are extremely
challenging. Therefore, the discussion of the nature of separate
tricritical points or a single tetracritical point with coil-globule
transition lines extending into the desorbed and the adsorbed
phases and the crystallization behavior near the adsorption line
should be left as a future work.

Finally, something should be said about the correction-to-scaling exponent
$\omega$. In all of the fits, we have noted that it did not change significantly
as we changed the parameter $s$, in contrast to $\phi$ and
$1/\delta$. In addition, the fits are not sensitive upon variation of 
$\omega$. Thus, all of our results have been obtained using an exponent
$\omega=0.5(1)$.

\section{Additional Comments and Concluding Remarks}

The adsorption/desorption transition of long polymers grafted on a surface has been
studied in simulations employing the contact-density
chain-growth algorithm, where the density of contacts is directly obtained.
This quantity can be used to analyze the thermodynamic behavior for any value of
temperature and solvent conditions. By using finite-size scaling theory,
taking into account properly the corrections to scaling, we constructed 
the phase diagram in the temperature versus solvent parameter space and estimated
the critical exponents. Our results are comparable to
estimates found in the literature for good solvent condition ($s=0$). In
particular, they agree very well with those reported by Klushin et al.~\cite{binder}, 
obtained by a different approach. On the
other hand, the agreement with the results from Luo~\cite{luo} is less
satisfactory. Although Luo has used a similar FSS for some quantities, the
present work has considered an additional scaling relation [given by Eq.~(\ref{dscal})] 
and resorted to corrections to scaling. The present values are also quite
comparable to the more recent ones obtained by Bradly et al. \cite{brad}, who
have likewise included corrections to scaling in their fits.

The phase diagram and the critical exponents suggest that the
critical line is not universal. Moreover, the exponents present a peak near the
region $s\sim1.5$, indicating the existence of a possible multicritical point.
The presence of this multicritical point can be associated to the
existence of different conformational phases of the polymer in the adsorbed phase,
with first-order transitions between them. At least one of these first-order transition 
lines will end up at the multicritical point. However, whether or not there is only one 
first-order line or several lines, is not clear at the moment. In addition, 
the rather strong variation of the critical exponents, as well as the
corresponding critical temperature near this region, can be the cause of the
difficulty encountered in getting the criticality of the model, even for $s=0$.
Naturally, more simulations in the ordered adsorbed region should be very
welcome to precisely determine the behavior of the transition lines close to
the multicritical point. 

We have also estimated the adsorbed transition temperature $T_a$ from the position
of the maximum values of $\Gamma$, as depicted in Figs. \ref{s0nu}(a) and \ref{gas-22}.
However, this quantity turned out to be less robust than the fourth-order Binder
cumulant $U_4$ and seems to have similar pitfalls like the specific heat based measurement \cite{kraw}.

Regarding the critical exponents $\phi$ and $1/\delta$, it is apparent from the scaling 
theory that they are independent exponents, as has been reported by Klushin et al. 
\cite{binder}. However, in the recent work by Bradly et al. \cite{brad}, it has been shown 
that, although independent, $\phi$ and $1/\delta$ have the same value, at least for $s=0$. 
Indeed, with an argument
similar to that used in Ref. \cite{brad}, one can show that they are identical for $s=0$. 
For instance, close to the adsorbing transition temperature the free energy of the
polymer chain can be given by Eq. (\ref{gp}). On the other hand, the energy, U, is 
given by
\begin{equation}
U={{\partial \left(\beta G\right)}\over{\partial \beta}}=-{{1}\over{k_BT^2}}
{{\partial \left(\beta G\right)}\over{\partial T}},
\label{E}
\end{equation}
where $\beta=1/k_BT$. Close to the transition temperature $T_a$, $\beta G=\beta_a G$ where
$\beta_a=1/k_BT_a$, leading to
\begin{equation}
U=-k_B{\beta_a}^3N^{1/\delta-1}{\cal G}^\prime(x),
\label{Es}
\end{equation}
where ${\cal G}^\prime(x)$ is the derivative of ${\cal G}(x)$ with respect to $x$. Nonetheless, 
for $s=0$, one also has from Eq. (\ref{en})
\begin{equation}
U={{\langle E_s\rangle}\over{N}}=-\epsilon\langle n_s\rangle=-\epsilon_sN^{\phi-1}f_{ns}(x).
\label{Es}
\end{equation}
From the above equations one concludes that $\phi=1/\delta$.

We can see from Table \ref{tab}
that the values of $\phi$ and $1/\delta$ are not equal, but close taking into
account the numerical error.
We would also like to emphasize that it is far more efficient to directly simulate 
self-avoiding walks instead of interacting self-avoiding walks, as has been done
here with sophisticated algorithmic efforts. In particular, 
simpler methodologies would enable simulations of much longer chains for the case $s=0$, and
in this case a better test for the equality of $\phi$ and $1/\delta$ would be achieved.
However, we believe that the polymer lengths we have considered in this paper are 
sufficiently long to allow for the quantitative discussion of the adsorption transition 
for all solvent conditions.

The above argument, however, holds only for $s=0$, where the internal energy and the order
parameter are related. Such relation does not hold as soon as $s\ne 0$. The behavior
of the critical exponents $\phi$ and $1/\delta$, as a function of $s$, depicted in Fig.
\ref{phinu} clearly shows that despite being different for general solvent conditions, 
they do meet at $s=0$. The inset in Fig. \ref{phinu} shows the difference of the
critical exponents $\Delta=\phi-1/\delta$ as a function of the solvent conditions. Assuming
their equality at $s=0$, from the data of Table \ref{tab} we estimate $\phi=1/\delta=0.485(7)$, 
in quite good  agreement with $\phi=1/\delta=0.484(4)$ from Ref. \cite{brad}.

Another important issue raised in Ref. \cite{brad} concerns the universality of this
system. The 
present results indicate a non-universal behavior as a function of $s$, while Bradly
et al. claim that the critical exponents should be universal. They reached this conclusion by
studying the self-avoiding trail in the cubic lattice, which presented the same critical
exponents as the self-avoiding walk. However, the self-avoiding trail does not seem to 
correspond to any value of $s$ in our simulations. It would be, however, desirable 
to simulate the self-avoiding trail with monomer-monomer interactions in order to see 
whether the corresponding exponents will differ or not from the good solvent condition. 

We believe that it is still premature to decisively claim non-universal behavior with
different values for both exponents. In order to seek for a universal behavior we could,
in addition of present analysis, consider
the same exponent $1/\delta$ along the $s$ line and determine the corresponding transition
temperature $T_a$ with this exponent, as shown in Figs. \ref{s0ta} and \ref{tcs-202}. 
However, in doing so, a
different $T_a$ is obtained with the corresponding critical exponent $\phi$ not only
diferent from $1/\delta$ but also $s$ dependent. 

Still within the scope of universality, since the exponents should be equal for
$s=0$, such behavior seems to violate both universality and weak universality hypotheses
(as is well known, in the weak universality the exponents vary but their ratio is constant). 
This kind of violation of both hypothesis has been recently reported for the ferromagnetic 
phase transition
of (Sm$_{1-y}$Nd{$_y$)$_{0.52}$Sr$_{0.48}$MnO$_3$ single crystals with ($0.5 \le y \le 1$),
where the magnetic  exponents vary with Nd concentration $y$ \cite{khan}. 
Our polymer adsorption model can be seen as a similar example of such
new scaling behavior, providing also a generic route leading to continuous variation of
critical exponents and multi-criticality. In fact, one of 
the open major problems is the precise identification of multicritical points 
and their physical understanding. This study requires an in-depth treatment 
of the different structural phases of the polymer and the corresponding 
transition lines between them in the 
adsorbed and desorbed phases, which is 
far from being easy. Whereas there has been a history of progress, none of 
the existing results can be considered conclusive. The most recent approach 
to use a generalized microcanonical statistical  analysis method reveals a 
multitude of features whose explanation requires additional careful analyses. 
This discussion, however, is beyond the scope of this manuscript and will be 
published once the analysis of the structural transitions has been completed.

\section{ACKNOWLEDGMENTS}

The authors would like to thank Prof.
Thomas Prellberg for stimulating discussions.
This work has been supported partially by CNPq (Conselho Nacional de
Desenvolvimento Cient\'ifico e Tecnol\'ogico, Brazil) under Grant No.
402091/2012-4 and by the NSF under Grant No. DMR-1463241. PHLM also 
acknowledges FAPEMAT (Funda\c c\~ao de Amparo \`a Pesquisa do Estado de 
Mato Grosso) for the support (Grant No. 219575/2015).


\end{document}